\documentclass{article}
\usepackage{psfig}
\begin{document}
\title{\vspace{-3cm}
\LARGE\bf On SUSY-QM, fractal strings and steps towards a proof of the
Riemann hypothesis}
\author{Carlos Castro$^1$, Alex Granik$^2$ and Jorge Mahecha$^3$ \\
{\small\em $^1$Center for Theoretical Studies of Physical
Systems,}\\ {\small\em Clark Atlanta University, Atlanta,
Georgia, USA}\\
\smallskip
{\small\em $^2$Physics Department, University of the
Pacific, Stockton, California, USA}\\
\smallskip
{\small\em $^3$Departamento de F\'{\i}sica, Universidad de
Antioquia, Medell\'{\i}n, Colombia}}
\date{\today}

\maketitle
 
\begin{abstract}

We present, using spectral analysis, a possible way to prove the Riemann's
hypothesis (RH) that the only zeroes of the Riemann zeta-function are of
the form $s=1/2+i\lambda_n$. A supersymmetric quantum mechanical model is
proposed as an alternative way to prove the Riemann's conjecture, inspired
in the Hilbert-Polya proposal; it uses an inverse eigenvalue approach
associated with a system of $p$-adic harmonic oscillators. An
interpretation of the Riemann's fundamental relation $Z(s) = Z(1-s)$ as a
duality relation, from one fractal string $L$ to another dual fractal
string $L'$ is proposed.

\end{abstract}
 
\section{\bf Introduction}
\label{sec:intro}

Riemann's outstanding hypothesis (RH) that the non-trivial complex zeroes
of the zeta-function $\zeta(s)$ must be of the form $s = 1/2 \pm
i\lambda_n$, remains one of the open problems in pure mathematics. The
zeta-function has a relation with the number of prime numbers less than a
given quantity and the zeroes of zeta are deeply connected with the
distribution of primes \cite{riemann}. The spectral properties of the
zeroes are associated with the random statistical fluctuations of the
energy levels (quantum chaos) of a classical chaotic system \cite{main}.
Montgomery \cite{montgomery} has shown that the two-level correlation
function of the distribution of the zeroes is the same expression obtained
by Dyson using random matrices techniques corresponding to a Gaussian
unitary ensemble. See also \cite{titchmash}, \cite{berry}, \cite{selvam}
and \cite{connes}. An extensive compilation on zeta related papers can be
found at \cite{watkins}.

One can consider a $p$-adic stochastic process having an underlying hidden
Parisi-Sourlas supersymmetry, as the effective motion of a particle in a
potential which can be expanded in terms of an infinite collection of
$p$-adic harmonic oscillators with fundamental (Wick-rotated imaginary)
frequencies $\omega_p = i\ln~p$ ($p$ is a prime) and whose harmonics are
$\omega_{p,n} = i\ln~p^n$ (See~\cite{cccastro}). This $p$-adic harmonic
oscillator potential allowed to determine a one-to-one correspondence
between the amplitudes of oscillations $a_n$ (and phases) with the
imaginary parts of the zeroes of zeta $\lambda_n$, after solving a inverse
eigenvalue problem.

Pitk\"anen \cite{pitkanen} proposed an strategy for proving the Riemann
hypothesis inspired by orthogonality relations between eigenfunctions of a
non-Hermitian operator that describes superconformal transformations. In
his approach the states orthogonal to a vacuum state correspond to the
zeroes of Riemann zeta. However, a proof was not given.

The contents of this work are the following. In section \ref{sec:susy} we
consider the SUSY-QM model approach to the Hilbert-Polya proposal. In
section \ref{sec:zeroes} we define a different operator than Pitk\"anen's,
expressed in terms of the Jacobi theta series, and we find the
orthogonality relations among its eigenfunctions, and finally we prove
some theorems and present a novel approach to prove the RH. In section
\ref{sec:string} a discussion of the fractal string construction, in
relation to the Riemann zeta-function, given by Lapidus and Frankenhuysen
and a possible generalization is presented. Finally, some concluding
remarks concerning the multifractal distribution of prime numbers found by
Wolf and the distribution of lengths of fractal strings are provided.

\section{\bf A supersymmetric potential}
\label{sec:susy}

The Hilbert-Polya idea to prove the RH is the following \cite{katz}. If the zeroes
of the Riemann zeta function are $s_n=1/2+i\lambda_n$, then it must exist a Hermitic
operator $\hat{T}$ such that the $s_n$ are complex eigenvalues of the operator
$1/2+i\hat{T}$, in other words, the real values $\lambda_n$ are eigenvalues of
$\hat{T}$. Here we propose a way to construct such operator by using SUSY-QM
arguments.
 
One of us~\cite{cccastro}, was able to consider a $p$-adic stochastic
process having an underlying hidden Parisi-Sourlas supersymmetry, as the
effective motion of a particle in a potential which can be expanded in
terms of an infinite collection of $p$-adic harmonic oscillators with
fundamental (Wick-rotated imaginary) frequencies $\omega_p = i\ln~p$ ($p$
is a prime) and whose harmonics are $\omega_{p,n} = i\ln~p^n$. This
$p$-adic harmonic oscillator potential allowed to determine a one-to-one
correspondence between the amplitudes of oscillations $a_n$ (and phases)
with the imaginary parts of the zeroes of zeta, $\lambda_n$, after solving
a inverse eigenvalue problem.

In SUSY-QM two iso-spectral operators $\hat{H}^{(+)}$ and $\hat{H}^{(+)}$
are defined in terms of the so called SUSY-QM potential. Here we use the
SUSY-QM model proposed in \cite{cccastro} based on the pioneering work of
B. Julia \cite{watkins}, where the zeta-function and its fermionic version
were related to the partition function of a system of $p$-adic oscillators
in thermal equilibrium at a temperature $T$. The fermionic zeta-function
has zeroes at the same positions of the ordinary Riemann function plus a
zero at $1/2+i0$, this zero is associated to the SUSY ground state. See
also the reference \cite{watkins}.

We propose an ansatz for the following antisymmetrized SUSY QM potential
(a ``$p$-adic Fourier expansion''):
\begin{equation}
\Phi(x; a,b) = \prod\limits_p\sum_j (a^{(p)}_j p^{jx} + b^{(p)}_j p^{-jx})
- \prod\limits_p\sum_j (a^{(p)}_j p^{-jx} + b^{(p)}_j p^{jx}),
\label{eq:SUSY}
\end{equation}
$p$ = primes, $j$ = naturals, $a\equiv\{a^{(p)}_j\}$ and
$b\equiv\{b^{(p)}_j\}$. This comes from the following SUSY Schr\"odinger
equation associated with the $\hat H^{(+)}$ Hamiltonian \cite{junker},
\begin{equation}
\left(\frac{\partial}{\partial x}+\Phi\right)
\left(-\frac{\partial}{\partial x}+\Phi\right)\psi_n^{(+)}(x)
=\lambda^{(+)}_n\psi_n^{(+)}(x),
\label{eq:SUSYH}
\end{equation}
where we set $\hbar = 2m = 1$. SUSY imposes that
$\Phi(x;a,b)$ is antisymmetric in $x$ so it must vanish at
the origin. Hence, $\Phi^2 (x; a,b)$ is an even function of
$x$ so the left/right turning points obey: $x^{(n)}_L$ =
$-x^{(n)}_R$ for all orbits, for each $n$ = $1$, $2$, ...
We define $x_n=x^{(n)}_R$.

The quantization conditions using the fermionic phase path integral
calculation (which is not the same as the WKB and sometimes is called the
Comtet, Bandrauk and Campbell formula \cite{junker}) are, set $\hbar$ =
$2m$ = $1$, so all quantities are written in dimensionless variables for
simplicity,
\begin{equation}
I_n (x_n; \lambda_n; a,b)\equiv
4\int_0^{x_n} dx \left[\lambda_n
- \Phi^2 (x; a,b)\right]^{1/2} = \pi n,
\label{eq:WKB}
\end{equation}
for $n$ = $1$, $2$, ... and $\lambda_n$ are imaginary parts
of the zeroes of zeta.

The second set of equations are given by the definition of
the turning points of the bound state orbits:
\begin{equation}
\Phi^2 (x_n;a,b) = \lambda_n;\ n =1,2, ...
\label{eq:turning}
\end{equation}

So, from the two sets of equations (\ref{eq:WKB}) and (\ref{eq:turning})
we get what we are looking for: Amplitudes of $p$-adic harmonic
oscillators, $(a,b)$, and (right) turning points $x_n$, depending on all
the $\lambda_n$.

It is very plausible that due to some hidden symmetry of the inverse
scattering problem there may be many solutions for the amplitudes of the
$p$-adic harmonic oscillators and turning points; i.e. many different SUSY
potentials $\Phi (x; a,b)$ do the job for many different sets ($a,b$) and
$x_n$. Numbers related by a symmetry which leaves fixed the eigenvalues of
the SUSY QM model = imaginary parts of the zeroes of zeta (fixed points).
We are not concerned with this case now only with proving that a solution
exists.

We emphasize that the integral quantization condition for the energy of
the orbitals of the SUSY QM model based on the fermionic phase space
approximation to the path integral Eq. (\ref{eq:WKB}) are only valid for
shape-invariant potentials $V_{\pm} (x)$ of the partner Hamiltonians
$H_{\pm}$. Only for those cases one has that the energy levels obtained
from this approximation happen to be exact. Since imposing
shape-invariance amount to an additional constraint one can just forget
about this restriction and write down the integral conditions based on a
fermionic path integral calculation that yields nice approximate results.
It is far easier to solve the inverse eigenvalue problem in this way than
to try to invert the Shr\"odinger equations! Therefore, since we are
working in the opposite direction, and we do not wish to impose an
impossible constraint on the SUSY potential, and we impose that the energy
levels are the imaginary parts of the zeroes, then one concludes that the
amplitudes of the $p$-adic harmonic oscillator SUSY QM potential, Eq.
(\ref{eq:SUSY}), $(a,b)$ are only approximate by our method. The numerical
results will be given in a forthcoming publication. In the next section we
propose an approach for a direct proof of the RH. It is based on the idea
of relating the non-trivial zeroes of the $\zeta$ with orthogonalities
between eigenfunctions of a conveniently chosen operator. See
\cite{pitkanen}, \cite{elimorzer} and \cite{sp1}.

\section{\bf A way to prove the Riemann conjecture}
\label{sec:zeroes}

The essence of our proposal is based in finding the appropriate $D_1$
operator
\begin{equation}
D_1 = - \frac{d}{d\ln t} + \frac{d V}{d\ln t} + k,
\label{eq:opD}
\end{equation}
whose eigenvalues $s$ are complex-valued, and its eigenfunctions are given
by
\begin{equation}
\psi_s (t) = t^{-s+k} e^{V(t)}.
\end{equation}
Notice that $D_1 $ is not self adjoint with eigenvalues given by complex
valued numbers $s$.

Also we define a partner operator of $D_1$ as follows,
\begin{equation}
D_2 = \frac{d}{d\ln t} + \frac{d V}{d\ln t} + k.
\label{eq:opD1}
\end{equation}
For this operator we have
\begin{equation}
D_2\psi_s (t) =(-s+2k+2t\frac{dV}{dt})\psi_s(t).
\end{equation}

The key of our approach relies in choosing the $V$ to be related to the
Bernoulli string spectral counting function, given by a Jacobi theta
series,
\begin{equation}
e^{2 V(t)} = \sum\limits_{n=1}^\infty e^{-\pi n^2 t^l}.
\label{eq:jacobitheta}
\end{equation}
The Jacobi's theta series is deeply connected to the statistics of
Brownian motion and integral representations of the Riemann zeta-function
\cite{biane}.

Then we are considering a family of $D_1$ operators, each characterized by
two real numbers $k$ and $l$ to be chosen conveniently. Notice also that
$D_1$ is invariant under scale transformations of $t$ and $F=e^V$, due to
$dV/(d\ln t)=d\ln F/(d\ln t)$ \cite{pitkanen}. The $D_1$ defined in
\cite{pitkanen} has $k=0$ and a different definition of $F$.

Let's recall the functional equation of the Riemann zeta-function
\cite{karatsuba},
\begin{equation}
Z(s)\equiv\pi^{-s/2}\Gamma\left(\frac{s}{2}\right)\zeta(s)=
\pi^{-(1-s)/2}\Gamma\left(\frac{1-s}{2}\right)\zeta(1-s)\equiv Z(1-s).
\label{eq:RieFund}
\end{equation}

We define the inner product as follows:
\begin{equation}
\langle f\vert g\rangle=\int\limits_0^\infty f^* g\frac{dt}{t}.
\label{eq:innerpr}
\end{equation}
With this definition, the inner product of two $D_1$ eigenfunctions is
\begin{equation}
\langle\psi_{s_1}\vert\psi_{s_2}\rangle=\alpha\int\limits_0^\infty
e^{2V}t^{-s_{12}+2k-1}dt=\frac{\alpha}{l}
Z\left[\frac{2}{l}(2k-s_{12})\right],
\label{eq:innerprpsi}
\end{equation}
where we have defined $s_{12}=s_1^*+s_2=x_1+x_2+i(y_2-y_1)$ and used the
expressions (\ref{eq:jacobitheta}) and (\ref{eq:RieFund}). $\alpha$ is a
constant to be conveniently chosen so that the inner product in the
critical domain is semi positive definite. The measure of integration
$d\ln t$ is also scale invariant. The integral is performed after the
change of variables $t^l=x$, which gives $dt/t=(1/l)dx/x$, and using the
result of equation (\ref{eq:RieFundRel}), given by Voronin and Karatsuba's
book \cite{karatsuba}.

We recall that $Z$ is the fundamental Riemann function, expressed in terms
of the Jacobi theta series, $\omega(x)=\sum_1^\infty\exp(-\pi n^2x)$ (see
Karatsuba and Voronin \cite{patterson}),
\begin{equation}
\begin{array}{rl}
\displaystyle\int\limits_0^\infty
\sum\limits_{n=1}^\infty
e^{-\pi n^2x}x^{s/2-1}dx& = \\
&\\
&\displaystyle = \int_0^\infty x^{s/2 - 1} \omega (x) dx\\
&\\
&\displaystyle ={1\over s(s-1)}+\int_1^\infty [x^{s/2-1} +
x^{(1-s)/2 - 1}] \omega (x) dx\\
&\\
&\displaystyle = Z(s) = Z(1-s).
\end{array}
\label{eq:RieFundRel}
\end{equation}
The right side is defined for all $s$, {\it i.e.\/}, this formula gives
the analytic continuation of the function $Z(s)$ onto the entire complex
$s$-plane \cite{patterson}.

Having\ the\ relation $(\alpha/l) Z[(2/l)(2k-s)] = (\alpha/l)Z[1-2/l(2k -s)]$, from
$Z[(2/l)(2k-s)]$ we can obtain the other expression simply by taking $k\to k-l/4$,
$l\to-l$ and $\alpha\to-\alpha$. If, and only if, $8k-4 = l$, then the above
discrete transformations become: $k\to1-k$, $l\to-l$ and $\alpha\to-\alpha$. The
importance of the relation $8k-4=l$ will be seen shortly.

From (\ref{eq:innerprpsi}) we obtain the ``norm'' of any state
characterized by $s=x+iy$,
\begin{equation}
\langle\psi_s\vert\psi_s\rangle=\frac{\alpha}{l}
Z\left[\frac{4}{l}(k-x)\right].
\label{eq:norm}
\end{equation}
The ``norm'' of the vectors is the same for all $s$ having the
same $x$. We will choose the domain of the values of $s$,
$1-s$ such that they fall inside the critical domain:
$0<{\cal R}e(s)<1$ and $0 < {\cal R}e(1-s)<1$. We will see
soon why this is crucial. Recall that at the boundaries, $s
= 0$ and $s = 1$ we have that $Z(s) = \infty$.

We exclude ${\cal R}e(s)=0$ and ${\cal R}e(s)=1$ because Vall\'ee de la
Poussin-Hadamard theorem says there are no zeroes of $\zeta$ at $x=0$ and
$x=1$ \cite{patterson}.

In particular, for the critical line $x=1/2$, the value of the ``norm'' is
$(\alpha/l) Z[2(2k-1)/l]$. Since we will choose that $l=4(2k-1)$, the
``norm'' becomes $(\alpha/l) Z(1/2)=(\alpha/l)(-3.97...)$, for all states
in the critical line, independently of the chosen value of $k$. This
forces the value of $\alpha$ to be negative, in the critical domain, and
for this reason we shall fix it to be equal to $-l$.

A recently published report by Elizalde, Moretti and Zerbini \cite{elimorzer}, which
contains some comments about the first version of this paper, considers in detail
the consequences of equation (\ref{eq:RieFundRel}). One of them is that equation
(\ref{eq:map}) lose its original meaning as a scalar product. Despite of this, we
will loosely referring this map as a scalar product. The Hilbert space inner product
property is not required so that the eigenvalues can be also negative. The states
have real norm squared, which need not however be positive definite. Hermiticity
requirement implies that the states are orthogonal to the reference state and
correspond to the zeros at the critical line. The problem is whether there could be
also zeros outside the critical line but inside the critical strip.

Also we must caution the reader that our arguments do not rely on the validity of
the zeta-function regularization procedure \cite{elizalde} nor the analytic
extension of the $\zeta$, which precludes a rigorous interpretation of the right
hand side of (\ref{eq:RieFundRel}) like a scalar product. We simply can replace the
expression ``scalar product of $\psi_{s_1}$ and $\psi_{s_2}$'' by the map $S$
defined as
\begin{equation}
\begin{array}{rccl}
S:& {\cal C}\otimes{\cal C}&\to&{\cal C}\\
 &&&\\
&(s_1,s_2)              &\mapsto&\displaystyle 
S(s_1,s_2)=-Z\left(\frac{s_1^*+s_2}{2}\right).
\end{array}
\label{eq:map}
\end{equation}
In other words, our arguments do not rely on an evaluation of the integral
$\langle\psi_{s_1}\vert\psi_{s_2}\rangle$, but only in the mapping
$S(s_1,s_2)$.

Now we describe our proposal to prove the RH. From now, we shall set $k=1$, $l=4$.

{\it Th. 1\/}. If $s=x+i\lambda_n$, where $0<x<1$ and $\lambda_n$ is
such that $\zeta(1/2+i\lambda_n)=0$, then the states $\psi_s$ and
$\psi_{1-s}$ are orthogonal.

\begin{figure}[h]
\centerline{\psfig{figure=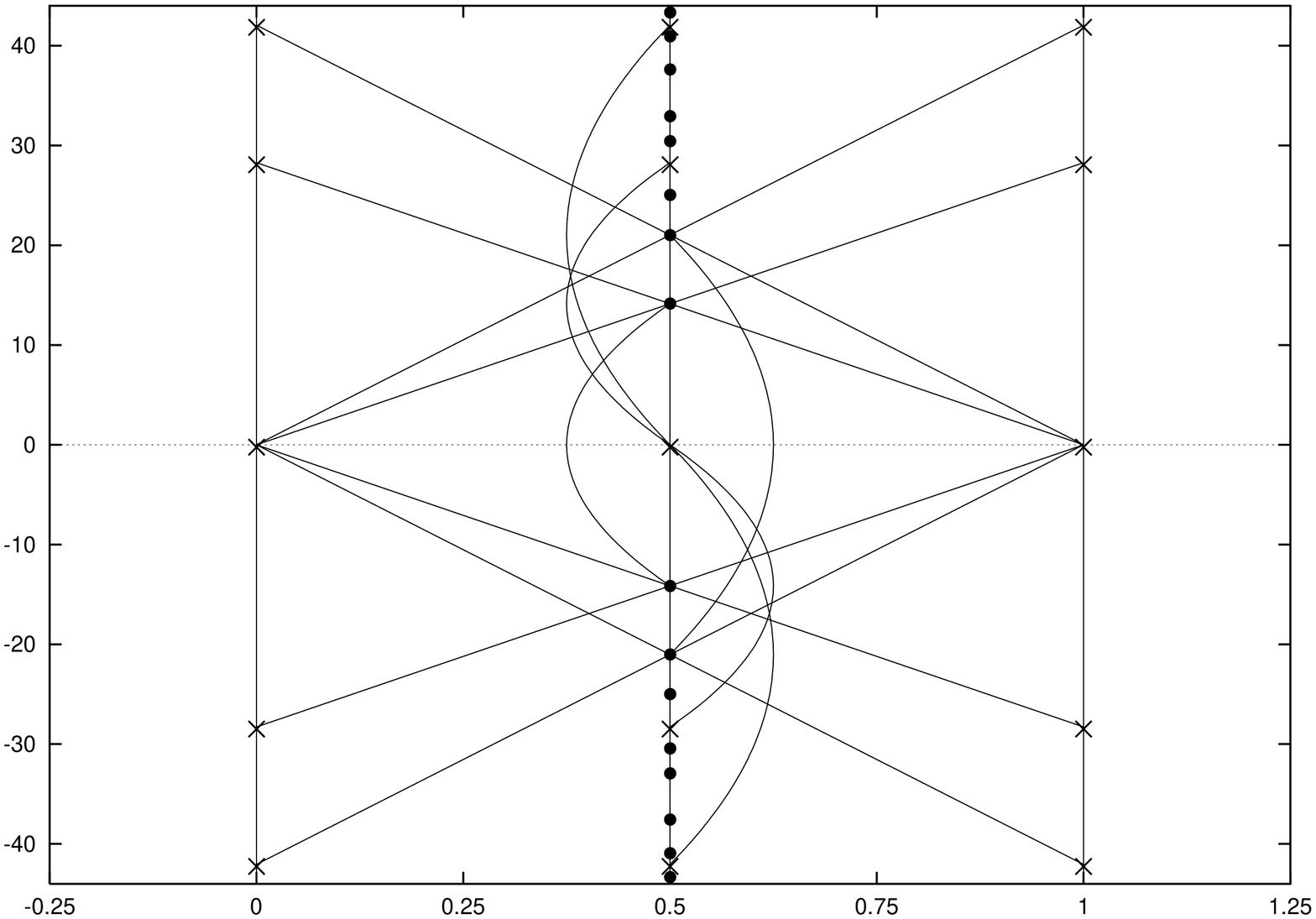,width=12cm,angle=270}}
\vspace{0.0mm} \caption[Short Title]{Lines between dots and crosses
represent pairwise orthogonal states. Dots represent zeroes of the
$\zeta$. Crosses on the critical line represent states orthogonal to the
$1/2+i0$. Crosses on the ${\cal R}e(s)=1$ represent states orthogonal to
the $0+i0$. Crosses on the ${\cal R}e(s)=0$ represent states orthogonal to
the $1+i0$. The states $1/2+i\lambda_n$ are orthogonal to
$1/2-i\lambda_n$. On the critical line, pairs of dots and/or pairs of
crosses are mutually orthogonal. Notice that for simplicity we are
representing the orthogonalities of states having only $x=0$, $x=1/2$ and
$x=1$. Here we are referring the particular case $k=1$, $l=4$.}
\label{fig:fig1}
\end{figure}

Proof: From (\ref{eq:innerprpsi}) it follows that
\begin{equation}
\langle\psi_s\vert\psi_{1-s}\rangle=
-Z(1/2+i\lambda_n)=0.
\label{eq:smenoss}
\end{equation}
Figure \ref{fig:fig1} represents those orthogonal states.

{\it Th. 2\/}. Any pair of states $s_1$ and $s_2$ symmetrically localized
with respect to the vertical line $x=1/2$, and such that
$y_1-y_2=2\lambda_n$, are orthogonal.

\begin{figure}[h]
\centerline{\psfig{figure=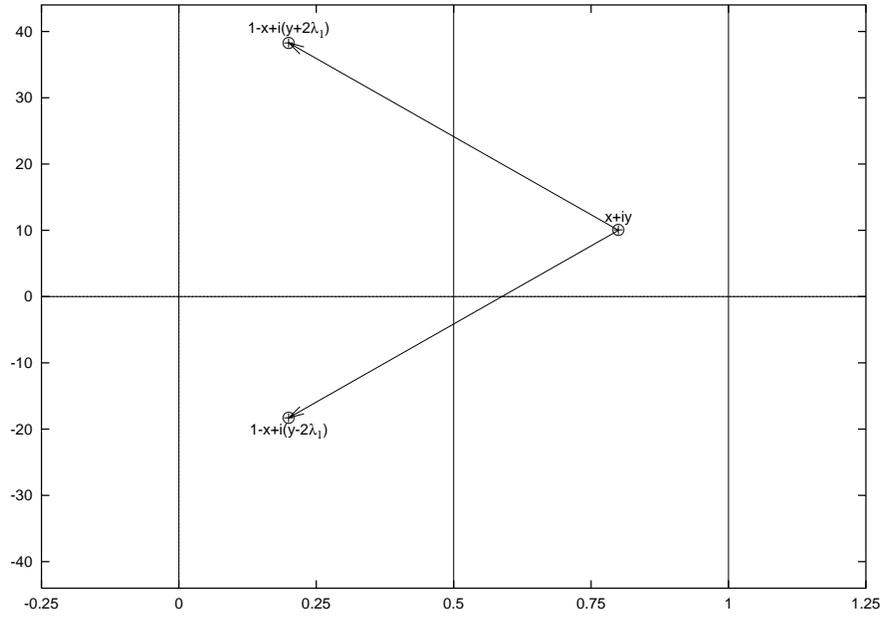,width=12cm,angle=270}}
\vspace{0.0mm}
\caption[Short Title]{ The state associated to a point $x+iy$ is
orthogonal to a doubly infinite series of states at
$1-x+i(y\pm2\lambda_n)$, for $n=1,2,... \infty$.
Here we are referring the particular case $k=1$, $l=4$.}
\label{fig:fig2}
\end{figure}

Proof: We can always find an $s=x+iy$ such that
$\langle\psi_{s_1}\vert\psi_{s_2}\rangle=
\langle\psi_s\vert\psi_{1-s}\rangle=0$. This follows straightforwardly, by
using (\ref{eq:innerprpsi}), $s_1^*+s_2=s^*+1-s=1-2iy$. Then, $x_1+x_2=1$
and $y_2-y_1=-2y$. By equating the arguments of the $Z$, it follows, from
theorem 1, that, $\langle\psi_s\vert\psi_{1-s}\rangle=0$ only if
$s=x+i\lambda_n$. Therefore, $y=\lambda_n$, $x_1+x_2=1$ and
$y_1-y_2=2\lambda_n$, as we intended to prove. Figure \ref{fig:fig2}
represents those states and we see that for any point $x+iy$ of the
complex plane within the critical strip, there is a doubly infinity family
of orthogonal states to it, given by $1-x+i(y\pm2\lambda_m)$ where
$\zeta(1/2+\lambda_m)=0$. As a special case we have that the orthogonal
states to the reference state $1/2+i0$ are located in the critical line.

{\it Th. 3\/}. The scalar product of two arbitrary states $s_1$
and $s_2$ within the critical strip is the same as the scalar product of a
third state $s_3$ of the critical strip with the reference state $1/2+i0$.

Proof: The statement follows directly from the definition
(\ref{eq:innerprpsi}) written for the pairs $s_1$ and $s_2$, and $1/2+i0$
and $s_3$. By using $\langle\psi_{s_1}\vert\psi_{s_2}\rangle=\
\langle\psi_{0+i0}\vert\psi_{s_1^*+s_2}\rangle=\
\langle\psi_{1/2+i0}\vert\psi_{s_1^*+s_2-1/2}\rangle$, yields directly
$s_3=s_1^*+s_2-1/2$.

Let's recall that $Z(s)=\pi^{-s/2}\Gamma(s/2)\zeta(s)$ and
$\langle\psi_{s_1}\vert\psi_{s_2}\rangle=-Z(1-s_1^*/2-s_2/2)$ for the
chosen values of $k$ and $l$. This scalar product can be rewritten as
$\langle\psi_{1/2+i0}\vert\psi_{s_1^*+s_2-1/2}\rangle$. If we define
$s=1-s_1^*/2-s_2/2$, so that $s_1^*+s_2-1/2=3/2-2s$, then the scalar
product between any states in the critical domain can be rewritten as,
\begin{equation}
\langle\psi_{s_1}\vert\psi_{s_2}\rangle=
\langle\psi_{1/2+i0}\vert\psi_{3/2-2s}\rangle=-Z(s).
\end{equation}
Due to the fact that $Z(s)=0$ if and only if $\zeta(s)=0$, therefore
$\zeta(s)=0$ if and only if $\psi_{3/2-2s}$ is orthogonal to
$\psi_{1/2+i0}$.
Then the RH is equivalent to the following statement: The orthogonal
states to the reference state are $1/2\pm2i\lambda_n$.

Now let's see an interesting consequence of theorem 1. If we define the
superposition
\begin{equation}
\vert\Psi\rangle = \vert\psi_s\rangle + \vert\psi_{1-s}\rangle,
\end{equation}
and evaluate the ``norm'' of $\Psi$, we note that the interference terms
are real valued
\begin{equation}
\langle\psi_s\vert\psi_{1-s}\rangle+
\langle\psi_{1-s}\vert\psi_s\rangle=
-Z\left(\frac{1}{2} + iy\right)
-Z\left(\frac{1}{2}- iy\right).
\label{eq:superp}
\end{equation}
The arguments of the $Z$'s are in the critical line. If $y=\lambda_n$, for
all these states, the norm-squared of the sum equals the sum of the
norm-squares. This is more precise, like the Pythagoras rule. See
\cite{pitkanen}. Therefore, the destructive interfering states are
distributed on infinite horizontal lines, each of them labelled by one of
the zeroes of the $\zeta$. The interference is destructive when $y$ in
$s=1/2+iy$ is such that $y=\pm\lambda_n$, with $\lambda_n$ one of the
zeroes of the $\zeta$.

Since on the critical line the norm-squared given by (\ref{eq:norm}) is
the same for all the states labeled by $1/2\pm iy$, we can visualize all
those states as living on a sphere and having the minimal ``norm''.

Notice the form of (\ref{eq:superp}). We get as we should a quantity plus
its complex conjugate for the interference which gives us a real valued
number. This makes perfect sense because the second inner product must be
the complex conjugate of the first inner product since we have reversed
the order of the inner product.

Summing up, the interference of the $\psi_s$ with $\psi_{1-s}$ is
destructive if and only if the values of $y_n = \lambda_n $. In such case
the norm-squared of the superposition equals the sum of the norm-squares
of its constituents, this is roughly speaking Pythagoras theorem. For
Pythagorean rational phases and the Riemann conjecture (see
\cite{pitkanen1}).

Now we explore some consequences of the mapping given by equation
(\ref{eq:map}). Let's suppose that $s=x+iy$ is a generic state orthogonal
to the reference state. Equation (\ref{eq:innerprpsi}) yields,
\begin{equation}
\langle\psi_{1/2+i0}\vert\psi_s\rangle=-Z(3/4-s/2)=-Z(1/4+s/2),
\end{equation}
together with their complex conjugates. It follows that if $x+iy$ is
orthogonal to $1/2+i0$, also orthogonal to this state are: $1-x+iy$,
$1-x-iy$ and $x-iy$.

The arguments of the $Z$ in each single one of those orthogonality
relations must be of the form $3/4-x/2\pm iy/2$ and $1/4+x/2\pm iy/2$.
From the definition of $Z$, there exists a zero if and only if
$\zeta(x'+iy')=0$. We are going to determine relations between $s'=x'+iy'$
and $s=x+iy$ such that the symmetries of the Z and the orthogonality
condition of states to the reference state are satisfied.

\begin{figure}[h]
\centerline{\psfig{figure=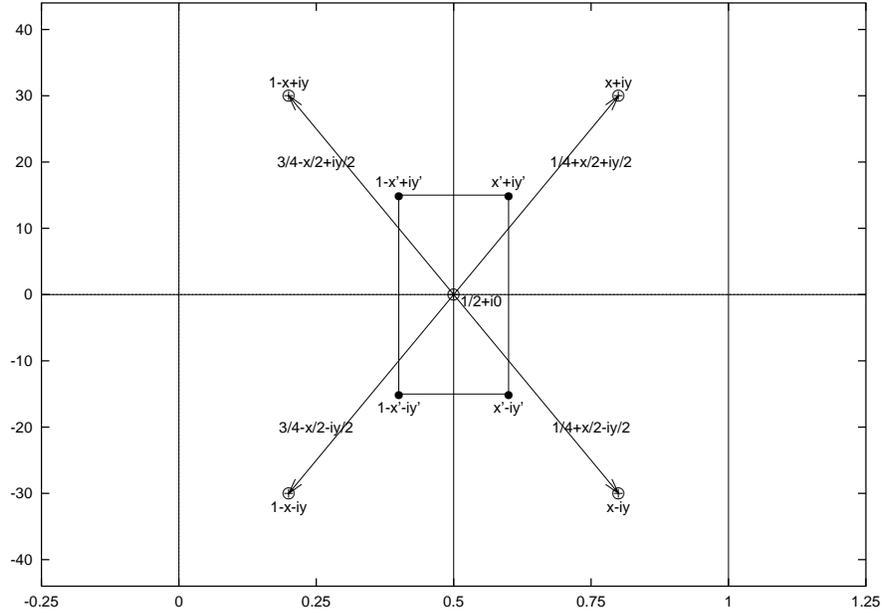,width=12cm,angle=270}}
\vspace{0.0mm} \caption[Short Title]{The dots represent generic zeroes of
the $\zeta$. The crosses represent generic states orthogonal to the
reference state $1/2+i0$. The numbers $3/4-x/2-iy/2$, etc, are the
arguments of $Z$ appearing in the orthogonality relations between states
orthogonal to the reference state. Due to the functional equation of the
Riemann zeta-function (\ref{eq:RieFund}), these arguments are just the
average values between $1/2+i0$ and those orthogonal states. Here we are
referring the particular case $k=1$, $l=4$.}
\label{fig:fig3}
\end{figure}

From the properties of the Riemann $\zeta$-function it follows that if
$s'$ is a zero, then there are three another zeroes, located at $s'^*$,
$1-s'$ and $1-s'^*$. See figure \ref{fig:fig3}. The 4 arguments of the $Z$
appearing in the orthogonality conditions must be related to the location
of the 4 zeroes of the $\zeta$. There are 24 orthogonality pairwise
combinations, easily identified if we maintain fixed the inner rectangle
of figure \ref{fig:fig3} and perform all permutations of the labels of the
orthogonal states.

If one studies all the possible families of mappings due to all the
possible $4!=24$ permutations of the vertices of the outer rectangle one
will inevitably introduce constraints among the four vertices of the same
rectangle; i. e. like $s^*=1-s$, and this will trivially lead to the RH.
This is unacceptable.

There are 4 relations of $s$ with $s'$ of the form $s=-1/2+2x'\pm2iy'$ or
$s=3/2-2x'\pm2iy'$, that we call generic cases. They are both
reflection-symmetric of each other. These simply state a correspondence
between the position of the orthogonal states and hypothetical zeroes
located in any point of the critical strip.

4 of the relations lead to an identification of the vertices of the
rectangle symmetrically located respect to the critical line, that is
$s'=1-s'^*$ or $s'^*=1-s'$, which implies the trivial result that those
vertices are located on the critical line and the imaginary parts are
related by $y=\pm2y'$. These 8 cases are compatible with the existence of
zeroes in the critical line.

The remaining 12 relations lead to an identification of all the four
vertices of the rectangle at the point $s'=1/2+i0$ which correspond to the
state $s=1/2+i0$. These cases do not correspond to any orthogonality
relation, due to in fact represent the inner product of the reference
state with itself.

It can not be discarded the presence of orthogonal states to the reference
located outside the critical line, the generic cases, and in consequence the
RH can not be deduced from this analysis.

Notice that the ``norm'' of the state corresponding to $s = 0$ is infinite
and that the states orthogonal to the $s = 0$ are those states whose $s =
1 + iy_n = 1 + 2i\lambda_n$. The steps to this proof of the RH did not
rely on the same {\it reductio ad absurdum\/} argument proposed by
Pitk\"anen (\cite{pitkanen}). In our approach, the RH is a consequence of
the symmetries of the orthogonal states as we intend to prove next, then
Th. 4 implies a proof of the RH assuming these symmetries.

{\it Th. 4\/}. The symmetries of the orthogonal states shown in figure 3
are preserved for any map $S$, equation (\ref{eq:map}), which give rise to
$Z(as+b)$, if $a$ and $b$ are such that $2b+a=1$.

Proof: If $s=x+iy$ is orthogonal to a reference state, then the Riemann zeta has
zeroes at $s'=x'+iy'$, $s'^*$, $1-s'$ and $1-s'^*$. If we equate $as+b=s'$, then
also $as^*+b=s'^*$. Now, $1-s'$ can be equated to $a(1-s)+b$ and $1-s'^*$ can be
equated to $a(1-s^*)+b$ if and only if if $2b+a=1$.

Notice that our election $a=-2/l$, $b=(4k-1)/l$ is compatible with this
symmetry if $k$ and $l$ are related by $l=4(2k-1)$. Reciprocally, if we
assume that the orthogonal states have the symmetries of figure 3, then
$a$ and $b$ must be related by $2b+a=1$, which gives rise to a very
specific relation between $k$ and $l$, obtained from $a+2b=1$; $a,b$ real.
The two generic cases correspond to taking $k=1$, $l=4$ and $k=0$, $l=-4$
respectively. It is clear that a map with arbitrary values of $a$ and $b$
does not preserve the mentioned symmetries. Theorem 4 contains a genuine
proof of the RH assuming that the invariance under the double symmetry is
true.

We have a family of $D_1$ operators, each labelled by $(k,l)$. Since we
can set $l=4(2k-1)$ due to the constraint $1=a(k,l)+2b(k,l)$ imposed by
the double-reflection symmetry, we can parametrize the eigenfunctions as
$\psi_s^{(k)}$, where $k$ and $s$ are continuous variables. Let's consider
two of those operators, $D_1^{(k_1)}$ and $D_1^{(k_2)}$. It must exist a
one to one correspondence between eigenfunctions of this pair of operators
at any given point $s$. We can see that to prove the RH is equivalent to
the following statement:

{\it St. (i)\/}. If $t=1/2+i0$ is the reference states and if $s_1$ and
$s_2$ are both points having $x_1=x_2=x$, then if $\psi_{s_1}^{(k_1)}$ is
orthogonal to $\psi_t^{(k_1)}$, then $\psi_{s_2}^{(k_2)}$ is also
orthogonal to $\psi_t^{(k_2)}$. In other words, the orthogonality of
states (with the same $x$) to the reference state is independent of $l$
and $k$.

Due to the fact that the inner products of the states of (i) with the
corresponding reference state are given by $Z[(2/l_1)(2k_1-t^*-s_1)]$ and
$Z[(2/l_2)(2k_2-t^*-s_2)]$, then the orthogonality gives rise to two
zeroes of the Riemann zeta. Those zeroes are of the form $s_1'=x'+iy_1'$
and $s_2'=x'+iy_2'$, that is, they are to be located on the same vertical
line.

This immediately allows us to write $s_1'=(2/l_1)(2k_1-t^*-s_1)$ and
$s_2'=(2/l_2)(2k_2-t^*-s_2)$, whose real parts are respectively
$x_1'=(2/l_1)(2k_1-1/2-x)$ and $x_2'=(2/l_2)(2k_2-1/2-x)$. If, and only
if, $x=1/2$ and $l_1=4(2k_1-1)$, $l_2=4(2k_2-1)$, then it follows that
$x_1'=x_2'=x=1/2$, $y_1'=(-2/l_1)y_1$, $y_2'=(-2/l_2)y_2$. Notice that the
values of $y_1$, $y_2$ do depend on $(k,l)$. $y_1'=\lambda_1$ and
$y_2'=\lambda_2$ are the imaginary parts of the two zeroes of $\zeta$.
Then (i) is equivalent to the RH.

These arguments can be generalized for all $(k,l)$ obeying $l=4(2k-1)$ as follows.
From $s'=a(k,l)s+b(k,l)$ and $a+2b=1$ we can write $s'=a(s-1/2)+1/2$. So
$x'=1/2+a(x-1/2)$ and, for a fixed $x$ the position of a zero could continuously
change by changing $(k,l)$. If the zeroes are supposed to form a discrete set, then
one must have that $x=1/2$, so that $x'=1/2$ is the only consistent value one can
have.

Another way of rephrasing this is saying that the family of the
$D_1^{(k,l)}$ operators yields a continuous family of pairs $(x,x')$
having the double-reflection symmetry $x\to1-x$, $x'\to1-x'$, from which
one arrives at the constraint $1/2=a(k,l)/2+b(k,l)$. This means that all
the curves given by $x'=a(k,l)x+b(k,l$ must contain the common point in
the $x-x'$ plane given by $(1/2,1/2)$, for all values of $(k,l)$, obeying
$l=4(2k-1)$. Also, if $\psi_s^{(k)}$ is orthogonal to the reference state
$1/2+i0$ then $s'$ is a zero of the $\zeta$, and the real parts are
related by $x'=a(x-1/2)+1/2$. Due to the statement (i) this real part must
to be independent of $k$ namely, independent of $a$. This can be satisfied
only if the orthogonal state satisfies $x=1/2$, which gives $x'=1/2$, the
RH.

Let us summarize what we have found so far, for $0<{\cal R}e(s)<1$. When
we choose $k=1$, $l=4$ then we have the following results (i), (ii),
(iii). (i) The interference between $\psi_s$ and $\psi_{1-s}$ is
destructive at the horizontal lines defined by $0<x<1$ and
$y=\pm\lambda_n$. (ii) If two states $s_1=1/2+i y_1$ and $s_2=1/2+i y_2$
on the critical line are orthogonal, then $y_1-y_2=\pm2\lambda_n$ for one
given $n$. (iii) Any pair of orthogonal states must obey $x_1+x_2=1$ and
$y_2-y_1=2\lambda_n$. And when we assume the double reflection symmetry,
then for all $(k,l)$ obeying $8k-4=l$ we have: (iv) The RH is a
consequence of the assumption that the orthogonal states to the reference
state have the same symmetry like the zeroes of $\zeta$ have.

\section{\bf The Riemann fractal string and its dual}
\label{sec:string}

Finally, we recall that the $e^{2V}$ defined by (\ref{eq:jacobitheta}),
expressed in terms of $x=t^4$, is nothing else than the energy partition
function of a Bernoulli string (whose standing waves have for wave vectors
integral multiples of the inverse of length \cite{lapidus}).

Several new relations of Riemann zeta-function with the spectral
properties of different physical systems have been found during the last
years. The authors discuss the Epstein zeta-functions \cite{lapidus} when
dealing with the Laplace operator in square/rectangular domains. The
counting function of the spectral eigenvalues of 2D-Laplacian gives the
Epstein zeta. Those functions also appear in the physics of $p$-branes
moving in hyperbolic spaces (negative constant curvature); i.e. in the
calculation of the effective potential of toroidal $p$-branes
living/embedded in target hyperbolic spaces \cite{neeman}

The spectrum of toroidal $p$-branes and branes moving in hyperbolic spaces
of constant negative curvature, is linked to the Epstein
zeta-functions~\cite{lapidus},
\begin{equation}
\zeta_{eps} (s)= \sum\limits_{\{n\}}\frac{1} {(n^2_1 + n^2_2 +...
n^2_{p+1})^{s/2}},
\end{equation} where $\{n\}\equiv\{n_1, n_2,... n_{p+1}\}$ is a non-zero
integer vector.

The Riemann zeta is a special case of the Epstein zeta and has a relation
to the spectrum of fractal strings (Alain Connes had urged to use a
different word to avoid confusion with string theory, suggests call them
``fractal harps'' instead). Two functions describe the fractal string, the
geometric length counting function $Z_L(s)$ and the frequency counting
function $Z_F(s)$ that are related by the fundamental relation involving
the Riemann zeta,
\begin{equation}
Z_F (s) = Z_L (s)\zeta (s).
\label{eq:LapRel}
\end{equation}

Let's imagine a chain made out of many links of sizes $l_1$, $l_2$, ...
$l_j$, ... The natural frequencies associated with the oscillations of
each one of those links are given by: $1/l_1$; $1/l_2$, ... $1/l_j$, ...
and the excitation states are just integer multiples of $1/l_j$: $k/l_j$;
for $k$ = $1$, $2$, ... j, ... There are described by the geometric length
counting function $Z_L(s)$ and the frequency counting function $Z_F (s)$.
The boundaries of the fractal strings are Cantor sets. See \cite{nasccast}
and \cite{nottale}.

Since we have selected the $V(t)$ of Eq.
(\ref{eq:jacobitheta}) to be related to the Bernoulli
string spectral counting function we must continue with the
fractal strings duality discussion. It is not a coincidence
that choosing $V(t)$ to be related to the Bernoulli string
was telling us something!

The length counting function is the sum,
\begin{equation}
Z_L(s) = \sum\limits_{j=1}^\infty l_j^s.
\end{equation}

The frequency counting function is the double sum,
\begin{equation}
Z_F(s) = \sum\limits_{k=1}^\infty\sum\limits_{j=1}^\infty k^{-s}l_j^s.
\end{equation}

And we always have $Z_F(s) = \zeta_{gen}(s) Z_L(s)$. In the case of a
fractal string it reduces to Eq. (\ref{eq:LapRel}). The book is entirely
based in choosing particular examples of fractals which fixed the lengths
$l_j$ {\it a priori\/} which enable to evaluate the sums explicitly. The
authors of \cite{lapidus} were able to find the {\it complex\/} dimensions
of the fractal strings by finding the location of the poles of the
$Z_L(s)$. It was essential that the poles do {\it not\/} coincide with the
zeroes of $\zeta(s)$.

Let us explain how one obtains the complex dimensions of the Cantor
string. The sequence of lengths must always be decreasing: $l_1$ $>$ $l_2$
$>$ ... or at most equal. Lets take the Cantor set as an example; there is
1 segment of length equal to 1/3; 2 segments of length = 1/9; ... $2^n$
segments of $l_n$ = $(1/3)^n$ and so forth. The degeneracy of each link
of length $l_n$ is $w_n$ = $2^n$. So the geometric counting function of
the Cantor string is:
\begin{equation}
Z_L(s)=\sum\limits_n (l_n)^s = \sum\limits_n
2^n\left(\frac{1}{3}\right)^{ns} = \frac{1}{1 - 2.3^{-s}}.
\end{equation}
so the poles of this sum yield the complex dimension: $1$ = $e^{i 2n \pi}$
= $2.(3)^{-s}$; taking the logarithm on both sides yields $s_n$ =
$(ln2/ln3)\pm i (2n\pi/\ln3)$ for the complex dimensions for $n$ = $0$,
$1$, $2$, ...

The real part is the standard dimension of the Cantor set. The imaginary
parts are periodic whose period is $2\pi/\ln3$.

We propose that the physical meaning of these complex valued dimensions
may be relevant for a theory of quantum gravity, because of interference
of complex dimensions.

It is very important to insist on the condition $0$ $<$
$\Re s$ $<$ $1$ since a fractal string of real valued
dimension $D$, embedded in a space of $R^d$, must satisfy
the constraint: $d - 1$ $<$ $D$ $<$ $d$.

The Cantor string, the golden string and the Fibonacci
string are defined respectively for the finite number of
scaling ratios given by
\begin{equation}
r_1 = r_2 = 1/3;\quad r_1 = 1/2, ~r_2 = 1/2^{1+\phi}, ~\phi
= 0.618...; r_1 = 1/2,~ r_2 = 1/3.
\end{equation}

We shall construct the Riemann fractal string (RFS) by different
procedures than the authors \cite{lapidus}, as the fractal boundaries of
the open 2-D domain.

The duality relation of the Riemann zeta $Z(s) = Z(1-s)$
only makes sense if, and only if, we write the continuum
limit for the length counting function
\begin{equation}
Z_L(s) = \int_0^\infty l(x)^s d \mu (x)\ {\rm and}\
Z_L(1-s) = \int_0^\infty \tilde{l}(x)^{1-s} d \tilde\mu
(x),
\label{eq:continuuml}
\end{equation}
and the frequency counting function
\begin{equation}
Z_F(s) = \int_0^\infty f(x)^{-s} d \mu (x)\ {\rm and}\
Z_F(1-s) = \int_0^\infty \tilde{f}(x)^{-1+s} d \tilde\mu
(x),
\label{eq:continuumf}
\end{equation}
where ($l(x)$, $\tilde{l}(x)$), ($f(x)$, $\tilde{f}(x)$), ($\mu (x)$,
$\tilde\mu(x)$) are the respective complex-valued, and their complex
conjugates, lengths, frequencies and measures associated with the
continuum limit of the RFS and their dual, that in the general case are
complex valued maps from the real line to the complex plane. The measure
$\mu(x)$ is such that $\vert \int_0^\infty d\mu(x)\vert = 1$.

In reference \cite{giona} a detailed analysis of the vector calculus and
contour integrals on fractal curves (boundaries of a bounded domain in the
complex plane) and interfaces is given. This was attained by constructing
pseudo-measures of integration based on iterated function systems.
Expressions for the contour integrals are obtained by means of a suitable
renormalization procedure (the length of a fractal contour is infinite)
and the solution of the Dirichlet problem on bounded two-dimensional
domains possessing fractal boundaries is given. These analytical tools
allow us to find in principle the complex valued functions appearing in
equation (\ref{eq:continuumz}) and to define the integrals in
(\ref{eq:continuuml},\ref{eq:continuumf}).

From (\ref{eq:continuuml}) and (\ref{eq:continuumf}) one can obtain the
generalization of the Lapidus and Frankenhuysen result, our proposal for
RFS,
\begin{equation}
\frac{Z_F}{Z_L}=\frac{\int_0^\infty f(x)^{-s} d \mu (x)}
{\int_0^\infty l(x)^s d \mu (x)} =\zeta(s),
\label{eq:continuumz}
\end{equation}
with analogous relation for the dual string. Lapidus and Frankenhuysen
results are recovered by using a Dirac delta distribution for the measure.

When $s$ is one of the zeroes of the $\zeta$, the equation
(\ref{eq:continuumz}) implies that the geometric length
counting function has a simple pole or that the frequency
counting function has a simple zero. On basis of this, we
define the Riemann fractal string as such string whose
geometric length counting function has a simple pole
precisely at the zeroes of the Riemann zeta-function.

\section{\bf Concluding remarks}
\label{sec:conclu}

In a previous work \cite{cj} we found very suggestive relations of the
golden mean and the distribution of the imaginary parts of the zeroes of
zeta. We pointed out that this could be related to the multifractal
character of the prime numbers distribution (See \cite{wolf}). Here we
observed yet another evidence of the multi-fractal distribution of primes
in the similarities in Wolf's formulae with those of Lapidus and
Frankenhuysen:
\begin{equation}
p(S_i) \sim l_i,\quad p(S_i)^q \sim l_i^s,\quad q \sim
s,\quad
\chi_q (l) \sim Z_L(s),\quad
l \sim L,
\end{equation}
where $L$ is the initial length where one begins to construct the fractal
string by defining $N$ finite segments of lengths $l_j=r_j L$,
$j=1,2,...N$ so that what is left is $L(1-R)$ where $R=\sum_1^N r_j < 1$.
One chooses $L$ so that $L(1-R)=1$ which means that the first length of the
iteration has unit length. From Wolf's paper we have that $p(S_i)$ is the
measure of the set $S_i$, the $\chi_q$ are the moments or partition
functions, which scale like a $l^{\tau(q)}$.

The Riemann zeta is deeply connected to fractal strings (whose boundaries
are Cantor sets) and it is no wonder why it is related to quantum chaos,
random matrix models, random walks, Brownian motion, etc.

\section*{\bf Acknowledgements} 

CC and JM acknowledge to the Center for Theoretical Studies of Physical Systems,
Clark Atlanta University, Atlanta, Georgia, USA, and the Research
Committee of the University of Antioquia (CODI), Medell\'{\i}n, Colombia
for support. Also very constructive comments from Matthew Watkins, School
of Mathematical Sciences University of Exeter, Matt Pitk\"anen from
Helsinki University and Gottfried Curio from Institute of Physics,
Humboldt University, are gratefully acknowledged. G. Curio found a
``vicious circle'' in our demonstration of the theorem 3 of the first
version of the paper. CC thanks M. Bowers and A. Boedo for their
hospitality in Santa Barbara, California where this work was initiated,
and L. A. Baquero for his hospitality in Medell\'{\i}n.

\newpage

\section*{\bf Figures} \label{sec:figs}
 
\def\baselinestretch{1}

\vfill
\newpage

\vfill
\newpage

\vfill
\newpage

\end{document}